\documentclass[a4paper,11pt]{article}
\pdfoutput=1 

\usepackage{jheppub} 

\usepackage[T1]{fontenc} 
\usepackage{graphicx}
\usepackage{hyperref}

\title{}

 \author[a,b]{Marco Bochicchio}

\affiliation[a]{Scuola Normale Superiore (SNS)\\Piazza dei Cavalieri 7, Pisa, I-56100, Italy}
\affiliation[b]{INFN sez. Roma 1\\Piazzale A. Moro 2, Roma, I-00185, Italy}

\emailAdd{marco.bochicchio@roma1.infn.it}

\abstract{We provide outstanding numerical evidence that in large-$N$ massless $QCD$ the joint spectrum of the masses squared, for fixed integer spin $s$ and unspecified parity and charge conjugation, obeys exactly the following laws:
$m_k^2=(k+\frac{s}{2}) \Lambda_{QCD}^2$ for $s$ even,  $m_k^2=2(k+\frac{s}{2}) \Lambda_{QCD}^2$ for $s$ odd,
$k=1,2,\cdots$ for glueballs, and $m_n^2=\frac{1}{2} (n+\frac{s}{2}) \Lambda_{QCD}^2$, $n=0,1,\cdots$ for mesons.
One of the striking features of these laws is that they imply that the glueball and meson masses squared form exactly-linear Regge trajectories in the large-$N$ limit of massless $QCD$,
all the way down to the low-lying states: A fact unsuspected so far. \par
The numerical evidence is based on lattice computations by Meyer-Teper in $SU(8)$ $YM$ for glueballs, and by Bali et al. in SU(17) quenched massless $QCD$ for mesons, that we analyze systematically.
The aforementioned spectrum for spin-$0$ glueballs is implied by a Topological Field Theory underlying the large-$N$ limit of $YM$, whose glueball propagators satisfy as well fundamental universal constraints arising from the asymptotic freedom and the renormalization group. No other presently existing model meets both the infrared spectrum and the ultraviolet constraints. We argue that some features of the aforementioned spectrum of glueballs and mesons of any spin could be explained by the existence of a Topological String Theory dual to the Topological Field Theory. }

\usepackage{braket}
\usepackage{amsmath}
\usepackage{amssymb}
\usepackage{graphicx}
\usepackage{hyperref}
\usepackage{fancyhdr}
\def\beq{\begin{equation}}
\def\eeq{\end{equation}}
\def\bea{\begin{eqnarray}}
\def\eea{\end{eqnarray}}
\def\bq{\begin{quote}}
\def\eq{\end{quote}}

\title{   Glueball and meson spectrum in large-$N$ massless $QCD$     }
\date{}
\begin{document}
\maketitle
%
%
%


\begin{figure}[t]
\centering
\includegraphics[width=.78\textwidth]{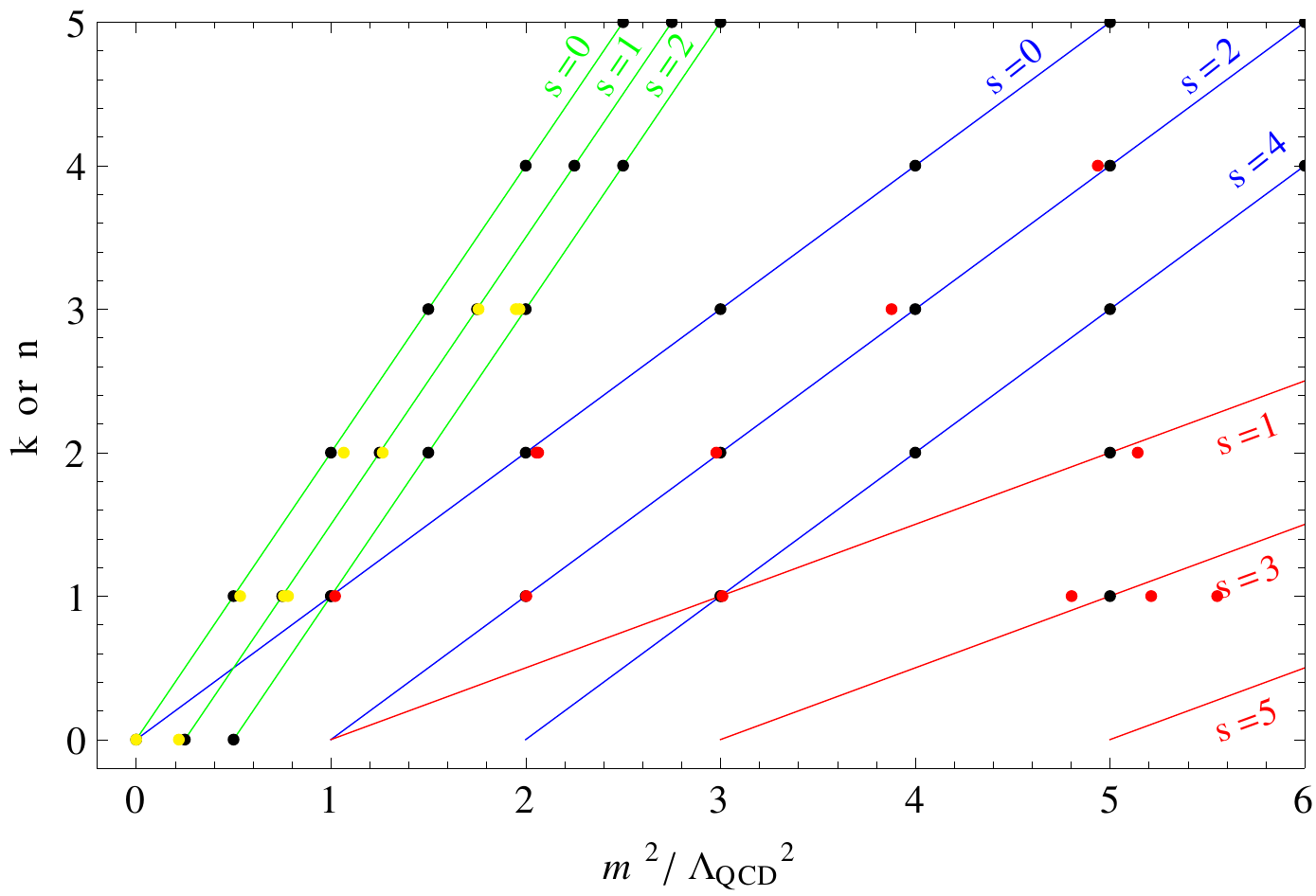}\\
\caption{The glueball and meson spectrum of large-$N$ massless $QCD$: The points in black, and the straight trajectories labelled by the spin $s$, represent the spectrum implied by the laws Eqs.~\eqref{eq:regge}. The red and yellow points represent
respectively glueballs and mesons actually found in the lattice computations~\cite{T,T0,L0,L}.}
\label{fig:regge1}
\vspace{3em}
\includegraphics[width=.78\textwidth]{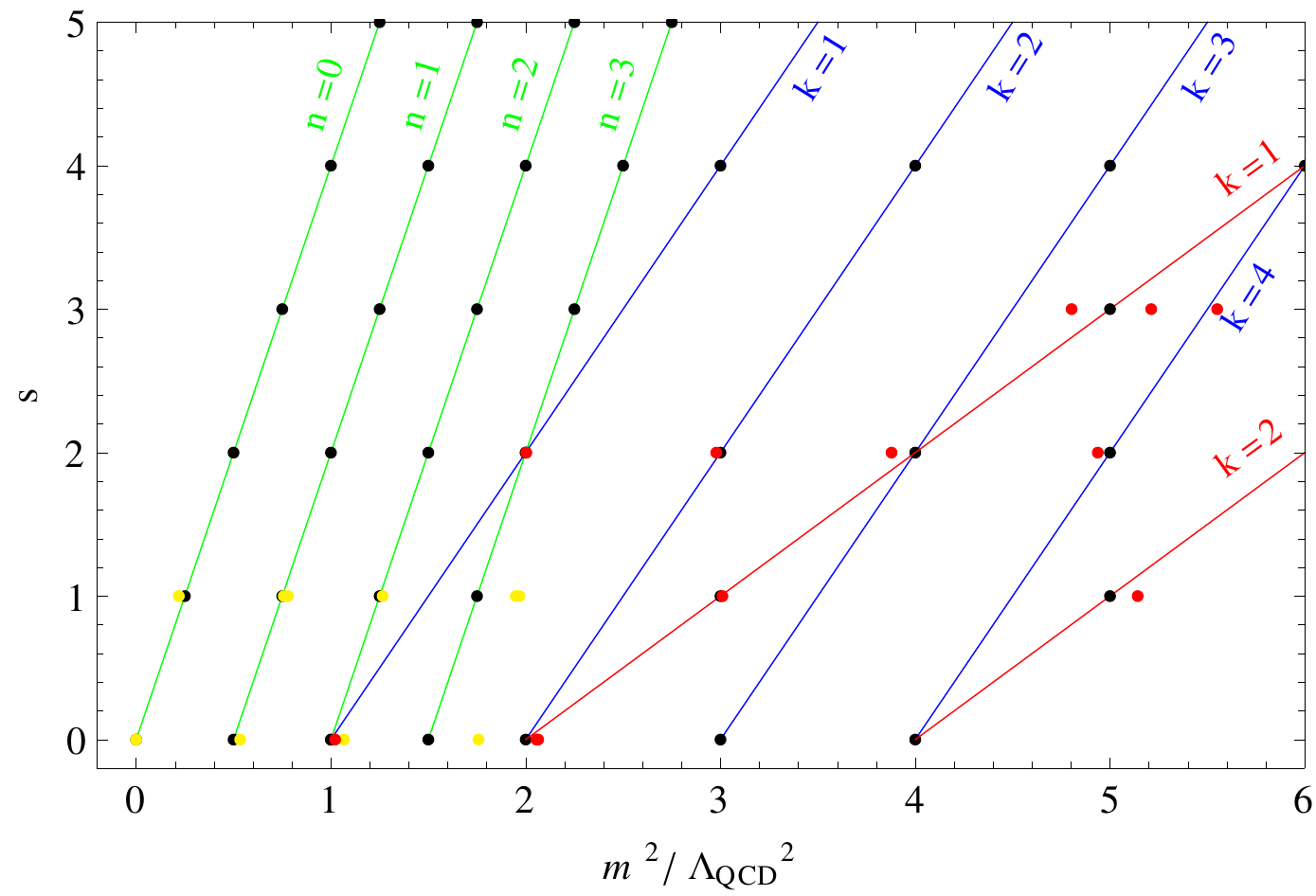}
\caption{The glueball and meson spectrum of large-$N$ massless $QCD$: The points in black, and the straight Regge trajectories labelled by the internal quantum numbers, $k$ for glueballs and $n$ for mesons, represent the spectrum implied by the laws Eqs.~\eqref{eq:regge}.}
\label{fig:regge2}
\end{figure}

\section{Introduction and Conclusions}

The main purpose of this paper is to provide outstanding numerical evidence that in `t~Hooft large-$N$ limit of massless $QCD$ \footnote{By massless $QCD$ we mean $QCD$ with massless quarks.} the glueball and meson spectrum for fixed integer spin $s$ obeys the following laws, exactly at the leading large-$N$ order. \par
For glueballs:
\begin{subequations}
\label{eq:regge}
 \begin{align}
  m_k^{(s)2}&= \left(k+\frac{s}{2}\right) \Lambda_{QCD}^2 \label{s1}\\
  \intertext{for $s$ even, and:}
  m_k^{(s)2}&= 2\left(k+\frac{s}{2}\right) \Lambda_{QCD}^2  \label{s2}\\
  \intertext{for $s$ odd, with $k=1,2,\cdots$ in both cases. For mesons:}
  m_n^{(s)2}&= \frac{1}{2} \left(n+\frac{s}{2}\right) \Lambda_{QCD}^2 \label{m}
 \end{align}
\end{subequations}
with $n=0,1,\cdots$. \par
$m_k^{(s)}$ and $m_n^{(s)}$ are the glueball and meson masses of spin $s$ respectively, and unspecified charge conjugation and parity, labelled by the internal quantum numbers $k$ and $n$ respectively, and $ \Lambda_{QCD}$ is the $QCD$ renormalization-group invariant scale in the scheme in which it coincides with the mass gap in the pure glue sector in `t~Hooft large-$N$ limit. \par
The numerical evidence is based on lattice computations by Meyer-Teper~\cite{T0,T} for glueballs and by Bali et al.~\cite{L0,L} for mesons, that we regard presently as the state of the art \footnote{Meyer-Teper computation for glueballs is presently on the largest lattice with the smallest value of the coupling and with the largest group $SU(8)$ in pure $YM$. Bali et al. computation for mesons involves presently the largest group $SU(17)$ in a carefully defined
large-$N$ massless limit of quenched lattice $QCD$.} of the subject of large-$N$ lattice $QCD$, and that we analyze systematically. \par
Phenomenologically, it is observed that the meson masses squared form Regge trajectories approximatively linear in the angular momentum. \par
Field theoretically, it is believed that the glueball and meson masses squared form Regge trajectories asymptotically linear in the angular momentum for large masses. \par
One of the striking features of Eqs.~\eqref{eq:regge} is that they imply that the glueball and meson masses squared in large-$N$ massless $QCD$ form exactly-linear Regge trajectories,
all the way down to the low-lying states: A fact unsuspected so far,
for which we provide outstanding numerical evidence in Section~\ref{2}. \par
Another striking feature is that Eqs.~\eqref{eq:regge} refer to the joint spectrum for fixed spin, without specifying parity $P$ and charge conjugation $C$. Thus the preceding equations do not identify uniquely any particular one-particle state occurring in the large-$N$ limit, but rather they describe the joint spectrum with respect to the other quantum
numbers $P$ and $C$, without specifying the degeneracy for fixed spin in the glueball and meson sectors. \par
 \par
Yet, since no numerical evidence can substitute a theoretical understanding, we report here below the theoretical reasons that made us to suspect that Eqs.~\eqref{eq:regge}
hold exactly in the large-$N$ limit. \par
Firstly, Eq.~\eqref{s1} restricted to the glueball sector with $s=0$, positive $C$ and unspecified $P$, holds exactly at the leading large-$N$ order in a Topological Field Theory ($TFT$) underlying the large-$N$ limit
of pure $YM$~\cite{MBN}. Obviously, the result extends to large-$N$ $QCD$ in `t~Hooft large-$N$ limit in the pure glue sector. \par
Roughly speaking the $TFT$ describes~\cite{MBN} the ground state of the large-$N$ one-loop integrable sector of massless $QCD$ of Ferretti-Heise-Zarembo~\cite{F}. 
The ground state in the pure glue sector is generated ~\cite{F} by scalar operators constructed by the anti-selfdual ($ASD$) curvature $F_{\alpha\beta}^-=F_{\alpha\beta}-\,^*\! F_{\alpha\beta}$ with $^*\! F_{\alpha\beta}=\frac{1}{2} \epsilon_{\alpha \beta \gamma \delta} F_{\gamma\delta}$. \par
Hence the $TFT$ computes~\cite{MBN} correlators that couple to scalar and pseudoscalar glueballs of positive $C$. The correlators in the $TFT$ factorize exactly~\cite{MBN} on the
spectrum in Eq.~\eqref{s1} for $s=0$, without being able to resolve the parity of the states.  \par
Besides, the chiral nature of the correlators in the $TFT$ suggested us that the proper generalization of the $s=0$ case should not specify other quantum numbers but the spin,
in order to fill all the points of the spectrum predicted by Eqs.~\eqref{eq:regge} without holes. Morally, this would occur should the points of the spectrum implied by Eqs.~\eqref{eq:regge} arise as poles in correlators
in the whole sector of Ferretti-Heise-Zarembo, including chiral fermion bilinears, that has a chiral nature~\cite{F} and that therefore cannot resolve the parity of the states that couple to the operators therein. \par
\begin{table}[t]
\begin{center}
\begin{tabular}{|c||c|c|c||c|c|}\hline
 & $m/\Lambda_{QCD}$ & $m/m_{0^{++}}$ & theor. & $\left(m/\Lambda_{QCD}\right)^2$ & theor.\\ \hline 
$0^{++}$ & $1.01$ & $1$ & $\sqrt{1}=1$ & $1.02$ & $1$ \\ \hline 
$0^{++*}$ & $1.43$ & $1.42$ & $\sqrt{2}=1.41$ & $2.06$ & $2$ \\ \hline 
$0^{-+}$ & $1.44$ & $1.42$ & $\sqrt{2}=1.41$ & $2.06$ & $2$ \\ \hline 
$2^{++}$ & $1.42$ & $1.40$ & $\sqrt{2}=1.41$ & $2.00$ & $2$ \\ \hline 
$2^{-+}$ & $1.73$ & $1.71$ & $\sqrt{3}=1.73$ & $2.98$ & $3$ \\ \hline 
$1^{+-}$ & $1.73$ & $1.72$ & $\sqrt{3}=1.73$ & $3.01$ & $3$ \\ \hline 
$2^{++*}$ & $1.97$ & $1.95$ & $\sqrt{4}=2$ & $3.88$ & $4$ \\ \hline 
$1^{--}$ & $2.27$ & $2.24$ & $\sqrt{5}=2.24$ & $5.14$ & $5$ \\ \hline 
$2^{--}$ & $2.22$ & $2.20$ & $\sqrt{5}=2.24$ & $4.94$ & $5$ \\ \hline 
$3^{++}$ & $2.19$ & $2.17$ & $\sqrt{5}=2.24$ & $4.80$ & $5$ \\ \hline 
$3^{--}$ & $2.28$ & $2.26$ & $\sqrt{5}=2.24$ & $5.21$ & $5$ \\ \hline 
$3^{+-}$ & $2.36$ & $2.33$ & $\sqrt{5}=2.24$ & $5.55$ & $5$ \\ \hline 
$?^{+-}$ & $2.40$ & $2.37$ & $?$ & $5.74$ & $?$ \\ \hline 
\end{tabular}
\end{center}
\caption{The glueball mass ratios implied by ref.~\cite{T} versus Eqs.~\eqref{eq:regge} (theor.). The $s=0$ case is actually a theoretical prediction of the $TFT$.}
\label{tab:glueballs}
\end{table}
A relevant feature of the $TFT$ is that it agrees sharply ~\cite{MBN} with Meyer-Teper numerical results for the lowest scalar and pseudoscalar states $[\frac{m_{0^{++*}}}{m_{0^{++}}}=1.419; \frac{m_{0^{-+}}}{m_{0^{++}}}=1.422]$ \footnote{We label glueballs and mesons by their $s^{PC}$ quantum numbers.}, since Eq.~\eqref{s1} predicts for the same ratios $\sqrt2=1.4142\cdots$.
Besides, the glueball correlators  of the $TFT$ agree \cite{MBN,MBM} crucially with fundamental universal constraints arising by the asymptotic freedom and the renormalization group, as opposed to the correlators computed by means of the $AdS$ String/Gauge Theory correspondence. Thus no presently existing model agrees both with the infrared and the ultraviolet of large-$N$ $QCD$ but the $TFT$~\cite{MBN}. For an analysis of all the implications of the $TFT$ versus
the $AdS$ String/Gauge Theory correspondence see ref.~\cite{MBN}. For an analysis of the ultraviolet asymptotics of glueball propagators in massless $QCD$ and in the $TFT$ see ref.~\cite{MBM}. For theoretical foundations of the $TFT$ see refs.~\cite{MB0,MB1,MBT}. For additional detailed computations see refs.~\cite{GGI,MBP}. \par
Secondly, were to exist a (Topological) String Theory $(TST)$ dual to the $TFT$, some features of the whole spectrum in  Eqs.~\eqref{eq:regge} could be understood as well as follows. \par
By construction, i.e. by the exact duality, the $TST$ would reproduce the $s=0$ case. \par
Moreover, it is natural to conjecture that the slopes, differing by a factor of $2$ between Eq.~\eqref{s1} and Eq.~\eqref{s2}, would correspond to the open \footnote{The $TST$ dual to the $TFT$ is essentially a theory of open strings, since the $TFT$ describes twistor Wilson loops
supported on Lagrangian submanifolds of twistor space of non-commutative space-time \cite{GGI}. Hence the dual topological string is the topological $A$-model on twistor space of non-commutative space-time, that describes open strings ending on Lagrangian submanifolds of twistor space of non-commutative space-time according to
section~14 of ref. \cite{GGI} and \cite{MBS}.}
and closed sectors of strings in the adjoint representation respectively, and that the further factor of $2$ for the slope in the meson sector Eq.~\eqref{m}, with respect to the would-be open  glueball sector in Eq.~\eqref{s1}, would correspond to an open string in the fundamental representation.\par
The specific form of Eqs.~\eqref{eq:regge} would be a dynamical feature of the
$TST$, but the integer or semi-integer character of the spectrum of the masses squared in units of a common scale squared should arise by its topological nature, since by construction the $TST$ would be the dual of the $TFT$
underlying $YM$. \par
We will work out elsewhere the spectrum of such $TST$, in order to check our stringy conjectures, along the lines foreseen in section 14 of ref.~\cite{GGI} and in \cite{MBS}. \par
Our conclusion, in addition to the main numerical evidence that is remarkable by itself, is that the aforementioned evidence favors the $TFT$ of $YM$ and indirectly the existence of a dual $TST$ as well, since no other presently existing model agrees~\cite{MBN} with Eq.~\eqref{s1} for $s=0$ in the infrared and with asymptotic freedom for glueball propagators in the ultraviolet, and globally with the actual glueball and meson spectrum as reported in Fig.~\ref{fig:regge1} and Fig.~\ref{fig:regge2}. \par
Moreover, we urge the lattice gauge theory community to further confirm or falsify Eqs.~\eqref{eq:regge}.

\section{Numerical Analysis} \label{2}
\begin{table}[t]
\begin{center}
\begin{tabular}{|c||c|c|c||c|c|}\hline
 & $m/\Lambda_{QCD}$ & $m/m_{0^{-+*}}$ & theor. & $\left(m/\Lambda_{QCD}\right)^2$ & theor. \\ \hline
$0^{-+}$ & $0$ & $0$ & $0$ & $0$ & $0$ \\ \hline 
$1^{--}$ & $0.468$ & $0.453$ & $\bigg.\sqrt{\tfrac{1}{4}}=0.500$ & $0.219$ & $\tfrac{1}{4}=0.25$ \\ \hline 
$0^{++}$ & $0.731$ & $0.708$ & $\bigg.\sqrt{\tfrac{1}{2}}=0.707$ & $0.534$ & $\tfrac{1}{2}=0.5$ \\ \hline 
$1^{++}$ & $0.871$ & $0.843$ & $\bigg.\sqrt{\tfrac{3}{4}}=0.866$ & $0.758$ & $\tfrac{3}{4}=0.75$ \\ \hline 
$1^{+-}$ & $0.883$ & $0.855$ & $\bigg.\sqrt{\tfrac{3}{4}}=0.866$ & $0.780$ & $\tfrac{3}{4}=0.75$ \\ \hline 
$0^{-+*}$ & $1.032$ & $1$ & $\sqrt{1}=1$ & $1.066$ & $1$ \\ \hline 
$1^{--*}$ & $1.125$ & $1.090$ & $\bigg.\sqrt{\tfrac{5}{4}}=1.118$ & $1.266$ & $\tfrac{5}{4}=1.25$ \\ \hline 
$0^{++*}$ & $1.326$ & $1.284$ & $\bigg.\sqrt{\tfrac{3}{2}}=1.225$ & $1.758$ & $\tfrac{3}{2}=1.5$ \\ \hline 
$1^{++*}$ & $1.396$ & $1.352$ & $\bigg.\sqrt{\tfrac{7}{4}}=1.323$ & $1.949$ & $\tfrac{7}{4}=1.75$ \\ \hline 
$1^{+-*}$ & $1.403$ & $1.359$ & $\bigg.\sqrt{\tfrac{7}{4}}=1.323$ & $1.968$ & $\tfrac{7}{4}=1.75$ \\ \hline  
\end{tabular}
\end{center}
\caption{The meson mass ratios implied by ref.~\cite{L0} versus Eqs.~\eqref{eq:regge} (theor.).}
\label{tab:mesons}
\end{table}
We start our numerical analysis by comparing the lowest mass glueball $k=1, s=0$ with the mass of the $n=2, s=0$ meson, in order to set a common $\Lambda_{QCD}$ scale for glueballs and mesons, since on the basis of Eqs.~\eqref{s1}--\eqref{m}
$m^{(0)}_{k=1}=m^{(0)}_{n=2}=\Lambda_{QCD}$. \par
The result of the lattice computations relevant for this paper were summarized by the authors of ref.~\cite{T} and of ref.~\cite{L0} in Tab.~\ref{tab:teper} and Tab.~\ref{tab:lucini} respectively, that we reproduce here. \par
In fact, we get numerically from Tab.~\ref{tab:teper} $[\frac{m^{(0)}_{k=1}}{\sqrt \sigma}]_{SU(8)}= 3.32 $~\cite{T} and from Tab.~\ref{tab:lucini} $[\frac{m^{(0)}_{n=2}}{\sqrt \sigma}]_{SU(\infty)}= 3.39 $~\cite{L0},
where $\sigma$ is the string tension measured in the lattice gauge theory, with the first ratio computed in $SU(8)$ and the second ratio extrapolated to $SU(\infty)$. \par
We assume that the first ratio is accurately close to its large-$N$ limit, as it is generally believed~\cite{LR}. Indeed, under this assumption,
the ratio $\frac{m^{(0)}_{n=2}}{m^{(0)}_{k=1}}=1.0211$ turns out to be $1$ with $2\%$ accuracy, consistently with Eqs.~\eqref{s1}--\eqref{m}. \par
\begin{table}[t]
\begin{center}
\begin{tabular}{|c|c|c|c|c|c|}
\hline
$J^{PC}$   & IR & $m/\sqrt{\sigma}$ &$\nu$ &  $\chi^2/(\nu-2)$&Average $m/\sqrt{\sigma}$\\
\hline
\hline
$0^{++}$      &$A_1$&  3.32(15)      & 4   &  0.41 & 3.32(15)\\
\hline
$0^{++*}$  &$A_1$&  4.71(29)      & 4   &  0.39 &  4.71(29)      \\
\hline
$2^{++}$  &$E$&  4.74(21)       & 4   &  0.20 &  4.65(19)   \\
  &$T_2$&  4.57(19)       & 4   &  0.45 &     \\
\hline
$2^{++*}$  &$E$&  6.47(50)      & 4   &  1.0 &    6.47(50)    \\
\hline
$3^{++}$  &$A_2$&  7.2(1.3)       & 3   &  0.08 &   7.2(1.3)  \\
\hline
\hline
$0^{-+}$      &$A_1$&  4.72(32)      & 4   &  1.1 &  4.72(32)   \\
\hline
 $?^{-+}$    &$T_1$&  7.87(77)      & 4   &  0.70 &  7.87(77) \\
\hline
$2^{-+}$      &$E$&  6.21(53)      & 4   &  0.28 & 5.67(40)\\
      &$T_2$&  5.36(40)      & 4   &  0.22 &  \\
\hline
\hline
$1^{+-}$   &$T_1$&  5.70(29)      & 4   &  0.85 & 5.70(29)  \\
\hline
$3^{+-}$   &$A_2$& 7.2(1.5)      & 3   &  0.09 & 7.74(79)  \\
           &$T_2$& 7.89(79)      & 3   &  0.18 &   \\
\hline
\hline
$1^{--}$   &$T_1$&  7.45(60)      & 4   &  0.07 & 7.45(60) \\
\hline
$2^{--}$      &$E$&  7.4(1.4)     & 3   &  0.87 &  7.3(1.4)\\
	      &$T_2$&  7.2(1.5)      & 3   &  0.01 &  \\
\hline
$3^{--}$   &$T_1$&  7.1(1.2)      & 3   &  0.001 &7.5(1.1)  \\
	   &$T_2$&  7.9(1.1)      & 3   &  0.004 &  \\
\hline
\hline

\end{tabular}
\end{center}
\caption{The $SU(8)$ glueball spectrum from ref.~\cite{T}.}
\label{tab:teper}
\end{table}
Besides, we fit by means of a global look at Fig.~\ref{fig:regge1} and Fig.~\ref{fig:regge2} $\Lambda_{QCD} \equiv \frac{(3.32)^2}{3.355} \sqrt \sigma $ and we compute all the mass ratios in units 
of $\Lambda_{QCD}$. We report the result for the ratios of glueball masses in Tab.~\ref{tab:glueballs}, and for the ratios of meson masses in Tab.~\ref{tab:mesons}.
The plots of the linear trajectories versus the observed states are displayed in Fig.~\ref{fig:regge1} and Fig.\ref{fig:regge2}. \par
For $10$ of the $12$ glueball states whose spin was identified in ref.~\cite{T} the actual difference between $\frac{m}{\Lambda_{QCD}}$ and the ratio implied by the laws in Eqs.~\eqref{eq:regge} is within about $1.5\%$ or better, for the $3^{++}$ state it is within about $2\%$ and for the remaining $3^{+-}$ state it is within $5\%$, but the $3^{+-}$ state has the largest estimated variance in Tab.~\ref{tab:glueballs}, so that a $5\%$ accuracy is still perfectly compatible. \par
In fact, the central values of the glueball mass ratios must be correlated much more than the estimate of their variance in ref.~\cite{T} implies, in order to explain such a good agreement
with Eqs.~\eqref{eq:regge}. \par
For the mesons the agreement with Eqs.~\eqref{eq:regge} is slightly worse: Only $3\%$ or better for $6$ of the $10$ states, including the pion that is exactly massless according to Eqs.~\eqref{eq:regge}, as it should be. \par
However, the meson masses are obtained numerically by extrapolating to the large-$N$ limit and to the massless limit but not to the continuum limit, that according to ref.~\cite{L0} may introduce additional systematic errors on the order of $5\%$. \par
The $\rho$ mass ratio agrees with Eqs.~\eqref{eq:regge} only within $7 \%$, but it may be more sensitive to the extrapolation to the massless limit, since the $\rho$ meson is the lightest massive meson.
By means of  Fig.~\ref{fig:regge1} and Tab.~\ref{tab:lucini} we observe that the $3$ most massive meson states would fit perfectly the Regge trajectory implied by Eqs.~\eqref{eq:regge}, were their spin shifted by $1$ with respect to the spin attributed in ref.~\cite{L0}. 
Otherwise, the difference of $\frac{m}{\Lambda_{QCD}}$ for the aforementioned $3$ meson states with respect to Eqs.~\eqref{eq:regge} is on the order of $6\%$ as it is seen in Fig~\ref{fig:regge2}, a fact still perfectly compatible with the $5\%$ possible estimated additional systematic error in ~\cite{L0}. In any case their masses squared are semi-integer valued in units of $\frac{1}{2}\Lambda^2_{QCD}$ within $2\%$ accuracy or better, a fact quite remarkable.\par
Taking the square doubles the errors, but the global agreement of $(\frac{m}{\Lambda_{QCD}})^2$ with Eqs.~\eqref{eq:regge} is still impressive, within about $3\%$ for $10$ of the $12$ glueball states and within about $6\%$ for $6$ of the $10$ meson states. 
Above all our numerical conclusions follow simply looking at these plots. \par

\begin{table}[t]
\begin{center}
\begin{tabular}[t]{|c|c|c|c|c|c|c|c|} 
\hline 
\cline{3-8} \multicolumn{2}{|c}{}  &\multicolumn{3}{|c|}{  ${m_\infty}/{ \sqrt{\sigma}}$}&\multicolumn{3}{|c|}{${m_\infty }/{F^\infty} $ }\\ \hline
Particle &$J^{PC}$&$m_q = 0$ & $m_q=m_{ud}$ &$m_q=m_{s}$ &$m_q = 0$ &$m_q=m_{ud}$&$m_q=m_{s}$\\ \hline
$\pi$ &$0^{-+}$  & 0 & 0.417(100) & 1.62(10) & 0 & 1.92(46) & 7.46(48) \\ \hline
$\rho$ &$1^{--}$ & 1.5382(65) & 1.6382(66) & 1.9130(79) & 7.08(10) & 7.54(11) & 8.80(13) \\ \hline
$a_0$ &$0^{++}$  & 2.401(31) & 2.493(31) & 2.755(32) & 11.04(21) & 11.47(22) & 12.67(23) \\ \hline
$a_1$ &$1^{++}$ & 2.860(21) & 2.938(21) & 3.158(22) & 13.16(21) & 13.51(21) & 14.53(23) \\ \hline
$b_1$ &$1^{+-}$  & 2.901(23) & 2.978(23) & 3.197(23) & 13.35(21) & 13.70(22) & 14.71(23) \\ \hline
$\pi^*$ &$0^{-+}$ & 3.392(57) & 3.462(57) & 3.659(58) & 15.61(34) & 15.93(35) & 16.83(36) \\ \hline
$\rho^*$ &$1^{--}$ & 3.696(54) & 3.756(54) & 3.928(54) & 17.00(34) & 17.28(35) & 18.07(36) \\ \hline
$a_0^*$ & $0^{++}$ & 4.356(65) & 4.420(65) & 4.603(66) & 20.04(41) & 20.33(41) & 21.18(42) \\ \hline
$a_1^*$ &$1^{++}$  & 4.587(75) & 4.646(75) & 4.816(77) & 21.10(46) & 21.38(46) & 22.15(47) \\ \hline
$b_1^*$ &$1^{+-}$ & 4.609(99) & 4.673(99) & 4.85(10) & 21.20(54) & 21.50(55) & 22.33(56) \\ \hline
$f_{\pi}^{\infty}$ & ---  & 0.3074(43) & 0.3271(44) & 0.3784(56) & $\sqrt{2}$ & 1.505(29) & 1.741(36) \\ \hline
$f_{\rho}^{\infty}$ & ---  & 0.5721(49) & 0.5855(50) & 0.6196(64) & 2.632(43) & 2.694(44) & 2.850(50) \\ \hline
\end{tabular}
\end{center}
\caption{The $N=\infty$ meson spectrum in quenched massless $QCD$
from ref.~\cite{L0}.}
\label{tab:lucini}
\end{table}

\section{Acknowledgements}

We would like to thank Barbara Mele and Giulio D'Agostini for a clarifying discussion on the variance of the mass ratios.
We would like to thank Alessandro Pilloni for working out the wonderful colorful plots in Fig.~\ref{fig:regge1} and Fig.~\ref{fig:regge2}.
 \thispagestyle{empty}

\end{document}